\documentstyle[preprint,aps]{revtex}

\begin{document}

\date{September 1994}
\preprint{LTP-037-UPR}
\title{Damping rate of a fermion in a medium}
\author{J. C. D'Olivo\thanks{
Partially supported by
Grant No. DGAPA-IN100691}}
\address{
	Instituto de Ciencias Nucleares\\
	Universidad Nacional Aut\'{o}noma de M\'{e}xico\\
	Apartado Postal 70-543, 04510 M\'{e}xico, D.F., M\'{e}xico
}
\author{Jos\'{e} F. Nieves\thanks{
Partially supported by the
US National Science Foundation Grant PHY-9320692}}
\address{
	Laboratory of Theoretical Physics\\
	Department of Physics, P. O. Box 23343\\
	University of Puerto Rico\\
R\'{\i}o Piedras, Puerto Rico 00931-3343
}


\maketitle

\begin{abstract}

We examine the relation between the damping rate
of a massless, chiral fermion that propagates
in a medium,
and the rate $\Gamma$ of approach to equilibrium.
It is proven that these quantities are equal, by
showing that they are given by the same formula
in terms of the imaginary part of the self-energy
evaluated at the energy of the propagating fermion mode.
This result is valid provided $\Gamma$ is defined
by  using the appropriate wave functions of the
mode.

\end{abstract}

It is well known that a massless fermion that
propagates through a medium acquires a dispersion
relation that differs from that in the vacuum\cite{weldon1}.
In general, the dispersion relation of a fermion
with momentum $p^\mu = (\varepsilon,\vec P)$ is not
given by $\varepsilon_P = P$ and, in particular,
$\varepsilon_P$ is not zero at zero momentum. Writing
\begin{equation}\label{eq1}
\varepsilon_P = \varepsilon_r - i\frac{\gamma}{2}\,,
\end{equation}
with $\varepsilon_r$ and $\gamma$ being real, the quantity
$M = \varepsilon_r(P = 0)$ can be interpreted as an effective mass,
and $\gamma$
as the fermion damping rate\cite{footnote1}.
In the framework of
Finite Temperature Field Theory (FTFT)\cite{holandeses,us1}, the quantity
of fundamental interest is the self-energy, from which
the dispersion relation is determined.

Besides the fermion damping rate,
a distinct concept, also related to the
imaginary part of the self-energy, has been used in the
literature.  This quantity, to which we will refer as
the total reaction rate, is denoted by $\Gamma$
and its inverse gives the time scale
for a distribution function to approach its
equilibrium value\cite{weldon2,kadanoff}.
In Ref.\cite{weldon2}, Weldon expressed $\Gamma$
as a combination of probabilities amplitudes
for various processess,
weighted by certain statistical factors
that account for the absortion and emission of the
particles by the medium.  By analysing the
one-loop contributions to boson and fermion
self-energies, he showed that $\Gamma$ could be
calculated from the imaginary  part of the self-energy
\cite{ashida}, although no attempt was made to connect
it  with the damping rate.

In this manner one ends up with two apparently different physical
quantities, the damping $\gamma$ and the total reaction rate
$\Gamma$, both related to the
imaginary part of the self-energy.
By studying the linear response
of a medium to an external field,
the equality between $\Gamma$ and $\gamma$ was established
for the collective color modes in a QCD plasma\cite{smilga}.
However, to our knowledge the same relation
has not been derived for the fermion case,
and this fact has been the source of some confussion.
In a recent article on the subject\cite{altherr},
two different formulas are used by the authors
to compute the damping and the total reaction
rate in terms of the self-energy
(Eqs. (2.2) and (2.27) of Ref. \cite{altherr}).
Our aim in this work is to show that
no such dichotomy is necessary and,
when correctly calculated in terms
of the fermion self-energy,  the damping and the  total
rate are given  by the same formula.
In this fashion, the result $\gamma = \Gamma$ is proved
to hold also for fermions.

Calculations of
$\gamma$  carried out during the last few
years within the realm of hot gauge theories,
have produced contradictory results,
entangled by questions of gauge invariance and other
problems\cite{pisarski}. It has been pointed out that
such ambiguities are a sympton of an incomplete
calculation, and a resummation of higher order corrections  is
necessary to get the correct result \cite{resummation}.
To avoid these complications
and make our presentation as clear as
possible, we examine here the simpler
but still instructive situation of fermions
interacting with a scalar field through a Yukawa interaction.

We begin by obtaining a formula for $\gamma$
in terms of the self-energy, which is valid  when
$\gamma\ll\varepsilon_r$\,.
This is the physically interesting regime
since otherwise the system would be overdamped
and the concept of a propagating mode
is not meaningful. That formula,
derived by expanding the equation for the dispersion
relation up to terms that are at most linear in
$\gamma$ and the absortive part of  the self-energy,
is the proper expression for the
fermion damping rate and it corrects or complements formulas that
have been quoted in previous works \cite{altherr,thoma}.
We then proceed to
demonstrate that the correct formula for $\gamma$
coincides with $\Gamma$, provided the latter quantity
is defined in terms of
the transition probabilities calculated with the
properly normalized spinors that satisfy the effective
Dirac equation in the medium instead of the free-particle
spinors.  The equality between the damping and the total
rate is first illustrated with a one-loop calculation,
and then a general proof is given using the
spectral representation of the fermion
self-energy.

%
%

The properties of a particle that
propagates through a medium are
determined by the linear part of the effective (classical)
field equation.  For a massless fermion with momentum $p^\mu$,
the equation, in momentum space, is
\begin{equation}\label{eq2}
(\rlap/ p - \Sigma_{eff})\psi = 0\,,
\end{equation}
with the self-energy $\Sigma_{eff}$ having the form
\begin{equation}\label{eq3.2}
\Sigma_{eff} = a\rlap/ p + b\rlap/ u\,,
\end{equation}
where $a$ and $b$ are functions of the variables
\begin{eqnarray}\label{eq3.3}
\varepsilon & = & p\cdot u\,,\nonumber\\
P & = & \sqrt{\varepsilon^2 - p^2}\,,
\end{eqnarray}
which give the energy and the magnitude of the 3-momentum
$\vec{P}$ in the rest frame of the medium.
We have introduced the vector $u^\mu$ representing the velocity
4-vector of the medium which in its own rest frame
has components $(1,\vec 0)$.

The self-energy can be decomposed
as
\begin{equation}\label{eq3.4}
\Sigma_{eff} = \Sigma_r + i\Sigma_i\,,
\end{equation}
in terms of the absorptive and dispersive parts
\begin{eqnarray}\label{eq3.5}
\Sigma_r & = & \frac{1}{2}(\Sigma_{eff} + \overline\Sigma_{eff})\,,\nonumber\\
\Sigma_i & = & \frac{1}{2i}(\Sigma_{eff} - \overline\Sigma_{eff})\,,
\end{eqnarray}
with
$
\overline\Sigma_{eff} = \gamma^0\Sigma_{eff}^\dagger\gamma^0
$.
Within the framework of the
real-time formulation of FTFT,
the dispersive part of the self-energy is given by
\begin{equation}\label{eq7}
\Sigma_r = \Sigma_{11r}\,,
\end{equation}
while the absorptive part can be calculated by any
of the following formulas
\begin{eqnarray}\label{eq8}
\Sigma_i & = & \frac{\Sigma_{11i}}{1 - 2n_F(x)}\nonumber\\
& = & \frac{i}{2}(\Sigma_{21} - \Sigma_{12})\nonumber\\
& = & \frac{\Sigma_{12}}{2in_F(x)}\,.
\end{eqnarray}
The $\Sigma_{ab}$ are the elements of
the $2\times 2$ self-energy
matrix to be computed using the Feynman rules of the theory.
In Eq.~(\ref{eq8}),
\begin{equation}\label{eq8a}
n_F(x) = \frac{1}{e^x + 1}
\end{equation}
is the Fermi-Dirac distribution for the fermion
that propagates in the medium,
written in terms of the variable
\begin{equation}\label{eq9}
x = \beta(p\cdot u - \mu)\,,
\end{equation}
where $\beta$ is the inverse temperature and $\mu$ is the chemical
potential of the fermion.

The dispersion relations of the particle and antiparticle are
determined by requiring Eq.~(\ref{eq2})
to have non-trivial solutions. The corresponding condition
determines also the poles of the fermion propagator and
can be written in the form
\begin{equation}\label{eq10}
f(\varepsilon,P)\overline f(\varepsilon,P) = 0\,,
\end{equation}
where
\begin{eqnarray}\label{eq11}
f(\varepsilon,P)\equiv (1 - a)(\varepsilon - P) - b\,,\nonumber\\
\overline f(\varepsilon,P)\equiv (1 - a)(\varepsilon + P) - b\,,
\end{eqnarray}
with $a$ and $b$ defined in Eq.~(\ref{eq3.2}).
Then, the dispersion relations $\varepsilon_P$ are obtained
as the solutions of
\begin{equation}\label{eq12}
f(\varepsilon_P,P) = 0
\end{equation}
and
\begin{equation}\label{eq13}
\overline f(\varepsilon_P,P) = 0\,.
\end{equation}
In general, the solutions $\varepsilon_P$ are complex, and a consistent
interpretation in terms of the dispersion relation for a mode propagating
through a medium is possible only if the imaginary part of
$\varepsilon_P$ is small compared to its real part.
In this
case the mode can be visualized as a particle with an energy and a
damping rate given by the real and imaginary part of $\varepsilon_P$,
respectively\cite{schrieffer}.

Under such assumptions, each one of
Eqs.~(\ref{eq12}) and (\ref{eq13})
yields two distinct solutions, one
with positive energy and the other with negative energy, whose
physical interpretation has been discussed in detail in Ref.~(\cite{weldon3}).
Here we will concentrate on the solution of Eq.~(\ref{eq12}) having
a positive real part, which corresponds to the particle
mode with energy $\varepsilon_r$, but
similar considerations and results apply to the other
solutions.

It is convenient to separate the function $f$ into
its dispersive and absorptive parts, and write\cite{footnote3}
\begin{equation}\label{eq14}
f(\varepsilon,P) = f_r(\varepsilon,P) + if_i(\varepsilon,P)\,,
\end{equation}
with a similar decomposition for $a$ and $b$.
Then
\begin{eqnarray}\label{eq15}
f_r & = & (1 - a_r)(\varepsilon - P) - b_r\,,\nonumber\\
f_i & = & -a_i(\varepsilon - P) - b_i.
\end{eqnarray}
Writing $\varepsilon_P$ in the form of Eq.~(\ref{eq1}),
the condition (\ref{eq12}) becomes
\begin{equation}\label{eq17}
f_r(\varepsilon_r - i\frac{\gamma}{2},P) +
if_i(\varepsilon_r - i\frac{\gamma}{2},P) = 0\,,
\end{equation}
which we solve by expanding in powers of $\gamma$ and retaining only terms
that are at most linear in $\gamma$ and $f_i$.  Thus, $\varepsilon_r$ is
determined as the solution of
\begin{equation}\label{eq18}
f_r(\varepsilon_r,P) = 0\,,
\end{equation}
while $\gamma$ is given by the formula
\begin{equation}\label{eq19}
\frac{\gamma}{2} = \frac{f_i(\varepsilon_r,P)}{N_P}\,,
\end{equation}
where
\begin{equation}\label{eq20}
N_P = \left[\frac{\partial f_r}
{\partial\varepsilon}\right]_{\varepsilon = \varepsilon_r}\,.\end{equation}
It is important to observe that the quantity $N_P$ coincides
with the normalization factor that has to be taken into account
in order to construct the spinors representing the one-particle states,
and which need to be included to correctly calculate the probability
amplitudes for the various processes involving the fermion\cite{us2}.
We also want to remark that our Eqs. (\ref{eq18}) and (\ref{eq19}),
with the factor $N_P$ included,
are analogous to those used in the context of
many-body physics\cite{manybody}.

The one-particle states have associated the wave functions $u_{L,R}$
satisfying the Dirac equation
\begin{equation}\label{eq21}
\left(\rlap/ p - \Sigma_r\right)u_{L,R} = 0\,,
\end{equation}
obtained by discarding the absorptive part of the self-energy.
While the explicit solutions of this equation are more easily worked
out in the rest frame of the medium, for
our present purposes the knowledge of the spinor projection operator
will be sufficient.  For the negative-helicity solution it is given by
\begin{equation}\label{eq22}
u_L\overline u_L = \frac{\varepsilon_r}{N_P}L\rlap/ n\,,
\end{equation}
where, as usual, $L = (1 - \gamma_5)/2$,
$N_P$ has been defined in Eq.~(\ref{eq20}) and
\begin{equation}\label{eq23}
n^\mu = \frac{1}{P}\left(p^\mu - (\varepsilon_r - P)u^\mu\right)
\end{equation}
is a null vector, whose components in the rest frame of the medium are
$n^\mu = (1,\frac{\vec P}{P})$.  There is a
relation for the positive-helicity solution $u_R$ similar
to Eq.~(\ref{eq22}).  For definiteness, in the rest of the article
we will work
with the negative-helicity solution,
but exactly the same considerations and results apply
to the other one as well.
Using
\begin{equation}\label{eq25}
\Sigma_i = a_i\rlap/ p + b_i\rlap/ u\,,
\end{equation}
and Eq.~(\ref{eq22}), it is easy to verify by direct
computation that the formula for the damping rate given in Eq.~(\ref{eq19})
can be written in the form
\begin{eqnarray}\label{eq24}
\gamma &  = & -\frac{1}{2N_P}\mbox{Tr}\, \left[
\rlap/ n\Sigma_i(\varepsilon_r,P) \right]
\nonumber\\
& = & -\frac{1}{\varepsilon_r}
\overline u_L\Sigma_i(\varepsilon_r,P)u_L\,.
\end{eqnarray}

In Eq.~(\ref{eq24}) we have indicated explicitly that $\Sigma_i$ must
be evaluated at $\varepsilon = \varepsilon_r$.
This a convenient formula for $\gamma$ that
will allow us to identify it with the
rate $\Gamma$ computed by squaring
amplitudes and integrating over the phase space with the appropriate
statistical weights.  One should become aware of
the general nature of Eq.~(\ref{eq24}),
which we have derived under the assumption that the damping is small,
but otherwise not tied in anyway to a perturbative, or for
that matter any other, particular method of
calculation of the self-energy.

If $\varepsilon_r$ and $N_P$ are approximated by their value in vacuum
($\varepsilon_r \simeq P$ and $N_P \simeq 1$),
then it is easy to see
from Eq.~(\ref{eq19}) that
$\gamma \simeq -2\mbox{Im}\,b(P,P)$, which coincides
with the formula given in Ref.~(\cite{thoma}) for the damping rate
in the high momentum limit ($P\gg T$).  However,
in other momentum regimes
\begin{displaymath}
\gamma\not = -2\mbox{Im}\,b(\varepsilon_r,P)\,,
\end{displaymath}
and the damping rate must be calculated as given in
Eq.~(\ref{eq19}) (or  Eq.~(\ref{eq24})), and by using the true dispersion
relation
in the medium $\varepsilon_r$.  In particular, the formula for
the damping rate quoted
in Eq. (2.27) of
Ref.~(\cite{altherr}), which is equivalent to put
$\gamma = -2\mbox{Im}\,b(\varepsilon_r,P)$, does not
correspond to the imaginary part of the energy momentum-relation
and does not coincide with the total reaction rate.

The next step is to establish
the relation between $\Sigma_i$
and the total reaction rate $\Gamma$,
which we will do first by using a 1-loop
example calculation of $\Sigma_i$.
As an instructive example, that is free of the
ambiguities associated with gauge invariance
pointed out in the beginning, we adopt the
model of a chiral fermion field $f_{L}$
interacting with a scalar field $\phi$ and a massive
fermion $\psi$ according to
\begin{equation}\label{eq25b}
L_{\rm int} = \lambda\overline \psi_R f_L\phi + h.c.
\end{equation}
The simplest way to proceed with the calculation is to
compute $\Sigma_{12}$ and then
use Eq.~(\ref{eq8}).
The relevant diagram is shown in Fig. 1 and,
according to the Feynman rules, its contribution is given by
\begin{equation}\label{eq26}
-i\Sigma_{12}(p) = (i\lambda)(-i\lambda)\int{\frac{d^4 p^\prime}{(2\pi)^4}
i\Delta_{21}(p^\prime - p)RiS_{12}(p^\prime)L\,.
}
\end{equation}
The components of the propagator matrices for the massive fermion and the
scalar
are
\begin{eqnarray}\label{eq27}
S_{12}(p^\prime) & = & 2\pi i\delta(p^{\prime 2} - m_\psi^2)
\left[\eta_F(p^\prime) - \theta(-p^\prime\cdot u)\right](\rlap/ p^\prime +
m_\psi)
\,,\nonumber\\
\Delta_{21}(k) & = & -2\pi i\delta(k^2 - m_\phi^2)
\left[\eta_B(k) + \theta(k\cdot u)\right]\,,
\end{eqnarray}
where, in terms of the variables
\begin{eqnarray}\label{eq28}
x^\prime & = & \beta(p^\prime\cdot u - \mu_\psi)\,,\nonumber\\
x_\phi & = & \beta(k\cdot u - \mu_\phi)\,,
\end{eqnarray}
we have
\begin{eqnarray}\label{eq29}
\eta_F(p^\prime) & = & \theta(p^\prime\cdot u)n_F(x^\prime) +
\theta(-p^\prime\cdot u)n_F(-x^\prime)\,,\nonumber\\
\eta_B(k) & = & \theta(k\cdot u)n_B(x_\phi) + \theta(-k\cdot u)n_B(-x_\phi)\,,
\end{eqnarray}
with $\theta$ being the step function.  The fermion distribution
$n_F$ is given in Eq.~(\ref{eq8a}), while for the scalar
\begin{equation}\label{eq30}
n_B(x_\phi) = \frac{1}{e^{x_\phi} - 1} .
\end{equation}
Momentum as well as
charge and lepton number conservation imply that
\begin{equation}\label{eq31}
x^\prime - x = x_\phi\,,
\end{equation}
where $x$ has been defined in Eq.~(\ref{eq9}).

The propagators $S_{12}$ and $\Delta_{21}$ can be rewritten using
the identities
\begin{eqnarray}\label{eq32}
\eta_F(p^\prime\cdot u) - \theta(-p^\prime\cdot u) & = & \epsilon(p^\prime\cdot
u)n_F(x^\prime)\,,\nonumber\\
\eta_B(k\cdot u) + \theta(k\cdot u) & = & \epsilon(k\cdot
u)n_B(x_\phi)e^{x_\phi}\,,
\end{eqnarray}
where $\epsilon(x) = \theta(x) - \theta(-x)$.
Substituting the resulting formulas into Eq.~(\ref{eq26})
and then using Eq.~(\ref{eq8}), we obtain
\begin{eqnarray}\label{eq33}
\Sigma_i & = & -\frac{\lambda^2}{8\pi^2}\int d^4 p^\prime\delta(p^{\prime 2} -
m_\psi^2)
\delta\left[(p^\prime - p)^2 - m_\phi^2\right]\nonumber\\
& & \mbox{} \times\epsilon(p^\prime\cdot u)
\epsilon\left((p^\prime - p)\cdot u\right)(n_F(x^\prime) + n_B(x_\phi))R(\rlap/
p^\prime + m_\psi)L .
\end{eqnarray}
To derive this result we used the equality
\begin{displaymath}
e^{x_\phi}n_F(x^\prime)n_B(x_\phi) = n_F(x)(n_F(x^\prime) + n_B(x_\phi))\,,
\end{displaymath}
which is a consequence of Eq.~(\ref{eq31}). The integral in Eq.~(\ref{eq33})
can be evaluated for different cases, but here we do not
go any further in that direction.  For our purposes it is more
convenient to express Eq.~(\ref{eq33}) as
\begin{eqnarray}\label{eq34}
\Sigma_i & = & -\frac{\lambda^2}{8\pi^2}\int{\frac{d^3
p^\prime}{2E^\prime}\frac{d^3 k}{2\omega_k}}
R\left[\delta^{(4)}(p + k - p^\prime)(n_\psi + n_\phi)(\rlap/ p^\prime +
m_\psi)\right.\nonumber\\
& & \mbox{} + \delta^{(4)}(p - k + p^\prime)(\overline n_\psi + \overline
n_\phi)(\rlap/ p^\prime - m_\psi)\nonumber\\
& & \mbox{} + \delta^{(4)}(p - k - p^\prime)(1 - n_\psi + \overline
n_\phi)(\rlap/ p^\prime + m_\psi)\nonumber\\
& & \mbox{} + \delta^{(4)}(p + k + p^\prime)(1 - \overline n_\psi +
n_\phi)(\rlap/ p^\prime - m_\psi)
\left.\right]L,
\end{eqnarray}
where $n_\psi$ and $n_\phi$ are the particle density distributions
\begin{eqnarray}\label{eq35}
n_\psi & = & \frac{1}{e^{\beta(p^\prime\cdot u - \mu_\psi)} + 1}\,,\nonumber\\
n_\phi & = & \frac{1}{e^{\beta(k\cdot u - \mu_\phi)} - 1}\,,
\end{eqnarray}
and $\overline n_{\psi,\phi}$ are the respective antiparticle
distributions,
obtained from $n_{\psi,\phi}$ by changing the sign of the chemical
potential.  In addition,
\begin{eqnarray*}
p^{\prime\mu} & = & (E^\prime,\vec p^{\,\prime})\,,\nonumber\\
k^\mu & = & (\omega_k,\vec k)\,,
\end{eqnarray*}
with
\begin{eqnarray*}
E^\prime & = & \sqrt{\vec p^{\,\prime 2} + m_\psi^2}\,,\nonumber\\
\omega_k & = & \sqrt{\vec k^2 + m_\phi^2}\,.
\end{eqnarray*}
Now, by means of the projection operators
\begin{eqnarray}\label{eq36}
\sum_{s}u_\psi(p^\prime,s)\overline u_\psi(p^\prime,s) = \rlap/ p^\prime +
m_\psi\,,\nonumber\\
\sum_{s}v_\psi(p^\prime,s)\overline v_\psi(p^\prime,s) = \rlap/ p^\prime -
m_\psi\,,
\end{eqnarray}
Eq.~(\ref{eq34}) leads to the relation
\begin{equation}\label{eq37}
\overline u_L\Sigma_i(\varepsilon_r,P)u_L = -\varepsilon_r\Gamma\,,
\end{equation}
where we have defined
\begin{eqnarray}\label{eq38}
\Gamma & \equiv & \frac{1}{2\varepsilon_r}
\int{\frac{d^3k}{(2\pi)^3 2\omega_k}\frac{d^3p^\prime}{(2\pi)^3
2E^\prime}}(2\pi)^4\nonumber\\
& & \mbox{}\times\left\{\right.\delta^{(4)}(p + k - p^\prime)[n_\psi(1 +
n_\phi) + n_\phi(1 - n_\psi)]{\sum_{s}|M_I|^2}\nonumber\\
& & \mbox{} + \delta^{(4)}(p - k + p^\prime)[\overline n_\psi(1 + \overline
n_\phi) + \overline n_\phi(1 - \overline
n_\psi)]{\sum_{s}|M_{II}|^2}\nonumber\\
& & \mbox{} + \delta^{(4)}(p - k - p^\prime)[(1 - n_\psi)(1 + \overline n_\phi)
+ n_\psi\overline n_\phi]{\sum_{s}|M_I|^2}\nonumber\\
& & \mbox{} + \delta^{(4)}(p + k + p^\prime)[(1 - \overline n_\psi)(1 + n_\phi)
+ \overline n_\psi n_\phi]{\sum_{s}|M_{II}|^2}
\left.\right\}
\end{eqnarray}
and
\begin{eqnarray}\label{eq39}
M_I & = & \lambda\overline u_\psi(p^\prime,s)u_L(p)\,,\nonumber\\
M_{II} & = & \lambda\overline v_\psi(p^\prime,s)u_L(p)\,.
\end{eqnarray}

The formula for $\Gamma$ given in Eq.~(\ref{eq38}) is immediately recognized
as the total rate for a particle of energy $\varepsilon_r$ and
momentum $P$ (as seen from the rest frame of the medium)
with integrations over the phase space weighted by
the statistical factors appropriate for each process.
$M_I$ is the
amplitude for $f\phi\rightarrow \psi$ or the decay
$f\rightarrow \psi\overline\phi$, while $M_{II}$ is the amplitude
for $f\overline\psi \rightarrow\phi$ or $f\phi\overline\psi\rightarrow 0$.
The amplitudes for the inverse reactions are given by the
complex conjugates of $M_I$ and $M_{II}$.  For specific values
of $\varepsilon_r$ and $P$, some of these processes will
be kinematically forbidden and will not contribute to $\Gamma$.
As is well known\cite{kadanoff2},
for Fermi systems the inverse reactions
are inhibited as a consequence of the Pauli blocking effect,
and they contribute additively to the depletion of the state.
Therefore, in this case $\Gamma$ is given by the sum of the rates
for the
direct and inverse processes, instead of their difference as in the
bosonic case.
Although our expression for $\Gamma$ in Eq.~(\ref{eq38}) has the
same form as the one used in Ref.\cite{weldon2},
it is important to keep in mind that
for us, the propagating
fermion is represented in the amplitudes $M_{I,II}$
by the spinor $u_L$ that obeys
the effective Dirac equation in the medium,
and not by a free particle spinor.
Comparing Eqs. (\ref{eq24}) and (\ref{eq37}) we finally obtain
the relation
\begin{equation}\label{eq40}
\gamma = \Gamma \,.
\end{equation}

%
%

We next show that the relation
between the matrix element of the absorptive part
$\Sigma_i$ of the self-energy and
the total rate, that was
established above by looking at the 1-loop calculation,
is a general result.
For this purpose we recall that,
in the real-time formulation of FTFT, the self-energy
of a fermion in a thermal background
is a $2\times 2$ matrix whose elements are defined by
\begin{eqnarray}\label{defsigma}
i\Sigma_{21}(z - y)_{\alpha\beta} & = & -\langle\eta_\alpha(z)
\overline\eta_\beta(y)\rangle\,,\nonumber\\
i\Sigma_{12}(z - y)_{\alpha\beta} & = & \langle\overline\eta_\beta(y)
\eta_\alpha(z)\rangle\,,\nonumber\\
-\Sigma_{11}(z - y) & = & \Sigma_{21}(z - y)\theta(z^0 - y^0)
+ \Sigma_{12}(z - y)\theta(y^0 - z^0)\,,\nonumber\\
-\Sigma_{22}(z - y) & = & \Sigma_{21}(z - y)\theta(y^0 - z^0)
+ \Sigma_{12}(z - y)\theta(z^0 - y^0)\,,
\end{eqnarray}
where $\eta$ and $\overline\eta$ are the fermion source
fields, in terms of which the interaction Lagrangian
is
\begin{equation}\label{Lint}
L_{\mbox{int}} = \overline f_L\eta + \overline\eta f_L\,.
\end{equation}
For the scalar model of
Eq.~(\ref{eq25b}) $\eta = \lambda\phi^\ast\psi_R$.
The angle
brackets in Eq.~(\ref{defsigma}) stand for the statistical average
which, for any operator ${\cal O}$, is defined by
\begin{equation}\label{defaverage}
\langle{\cal O}\rangle = \frac{1}{Z}\sum_i\langle i|\rho{\cal O}|i\rangle\,,
\end{equation}
where
\begin{equation}\label{rho}
\rho = e^{-\beta H + \sum_{A}\alpha_A Q_A}
\end{equation}
and
\begin{equation}\label{zeta}
Z = \sum_i\langle i|\rho|i\rangle\,,
\end{equation}
with $H$ being the Hamiltonian
of the system. The quantities $Q_A$ are the (conserved) charges
that commute with $H$, and the $\alpha_A$ are the chemical
potentials that characterize the composition of the background.

Using the convention of Eq.~(\ref{Lint}), the amplitude
for the decay $n + f(p)\rightarrow m$,
where $n$ and $m$ label any two states of the system,
is
\begin{equation}
A = (2\pi)^4\delta^{(4)}(p + q_n - q_m)
\langle m|\overline\eta(0)u_L(p)|n\rangle\,,
\end{equation}
where $u_L(p)$ is the properly normalized spinor
of the propagating mode $f$.  From this follows
that the total decay rate, averaged over the initial
states of the system, is
\begin{equation}
\Gamma_D = \frac{1}{Z}\frac{1}{2p^0}\sum_{n,m}
|\langle m|\overline\eta(0)u_L(p)|n\rangle|^2(2\pi)^4
\delta^{(4)}(p + q_n - q_m)Z_n\,,
\end{equation}
where $Z$ is defined in Eq.~(\ref{zeta}) and
\begin{eqnarray}
Z_n  & = & \langle n|\rho|n\rangle\nonumber\\
& = & e^{-\beta q_n\cdot u + \alpha_n}\,.
\end{eqnarray}
Here $\alpha_n$ is the eigenvalue of the operator
$\hat{\alpha} = \sum_A\alpha_A Q_A$ corresponding to
the state $|n\rangle$; i.e., $\hat{\alpha}|n\rangle =
\alpha_n|n\rangle$.  In similar fashion,
the total probability for the inverse decay
$m\rightarrow n + f(p)$ is given by
\begin{equation}
\int\frac{d^3p}{(2\pi)^3}\Gamma_I\,,
\end{equation}
where
\begin{equation}\label{GammaInv}
\Gamma_I = \frac{1}{Z}\frac{1}{2p^0}\sum_{n,m}
|\langle n|\overline u_L(p)\eta(0)|m\rangle|^2(2\pi)^4
\delta^{(4)}(p + q_n - q_m)Z_m\,.
\end{equation}
Since $\eta$ and $f$
have the same quantum numbers, then
\begin{displaymath}
\langle n|\overline u_L(p)\eta(0)|m\rangle \not = 0
\end{displaymath}
only for those states such that $\alpha_m =
\alpha_n + \alpha_f\,$.
This relation, combined with the momentum conservation
condition implied by the delta function
in the formulas for $\Gamma_{D,I}$ above, allows us
to replace in Eq.~(\ref{GammaInv})
\begin{equation}
Z_m = e^{-\beta p\cdot u + \alpha_f}Z_n\,,
\end{equation}
which gives
\begin{equation}\label{GammaID}
\Gamma_I = e^{-x}\Gamma_D\,,
\end{equation}
or equivalently
\begin{equation}
(1 - n_F(x))\Gamma_I = n_F(x)\Gamma_D\,,
\end{equation}
where $n_F$ and $x$ are given in Eqs.~(\ref{eq8a}) and
(\ref{eq9}), with $\mu = \alpha_f\beta$.

In order to establish the relation between $\Sigma_i$
and the total rate, we start from the defining Eq.~(\ref{defsigma}).
Inserting a complete set of states $\sum|n\rangle\langle n|$
between the
fields $\overline\eta(y)$ and $\eta(z)$, we have
\begin{eqnarray}\label{Sigma12specrep}
i\Sigma_{12}(z - y)_{\alpha\beta} & = & \frac{1}{Z}\sum_{n,m}e^{-i(q_m -
q_n)\cdot
(z - y)}\times\nonumber\\
& & \langle m|\overline\eta_\beta(0)|n\rangle\langle n|\eta_\alpha(0)
|m\rangle Z_m\,,
\end{eqnarray}
from which we immediately obtain
\begin{equation}\label{Sigma12GammaI}
i\overline u_L(p)\Sigma_{12}(p)u_L(p) = 2p^0\Gamma_I\,.
\end{equation}
In similar fashion,
\begin{eqnarray}\label{Sigma21specrep}
i\Sigma_{21}(z - y)_{\alpha\beta} & = & \frac{-1}{Z}\sum_{n,m}e^{-i(q_m -
q_n)\cdot
(z - y)}\times\nonumber\\
& & \langle n|\eta_\alpha(0)|m\rangle\langle m|\overline\eta_\beta(0)
|n\rangle Z_n\,,
\end{eqnarray}
so that
\begin{equation}
i\overline u_L(p)\Sigma_{21}(p)u_L(p) = -2p^0\Gamma_D\,.
\end{equation}
Using these results in the relation
$\Sigma_i = \frac{i}{2}(\Sigma_{21} - \Sigma_{12})$
[Eq.~(\ref{eq8})]
we finally obtain\cite{footnote4}
\begin{equation}\label{Sigmaitotalrate}
\overline u_L(p)\Sigma_i(p)u_L(p) = -p^0\Gamma\,,
\end{equation}
where
\begin{equation}\label{totalrate}
\Gamma = \Gamma_I + \Gamma_D\,.
\end{equation}
On the other hand, recall from
Eq.~(\ref{eq24})
that the damping rate, determined
as the imaginary part of the dispersion relation,
is given by
\begin{displaymath}
\gamma = -\frac{1}{\varepsilon_r}
\overline u_L\Sigma_i(\varepsilon_r,P)u_L\,.
\end{displaymath}
where $\varepsilon_r$ is the (real part of the) true
dispersion relation
of the propagating mode, and
$u_L$ is the corresponding spinor wave function.
With the spinor $u_L$ thus chosen and with $p^0 = \varepsilon_r$,
$\Gamma$ in Eq.~(\ref{totalrate})
has the interpretation of the total rate for all the processes
involving the propagating fermion,
so that Eqs.~(\ref{Sigmaitotalrate})
and (\ref{eq24})
establish the equality $\gamma = \Gamma$ with generality.

In conclussion,
the damping rate, determined from the imaginary part
of the dispersion relation, coincides with the total rate,
provided the latter is
calculated with the correct wave function of the propagating
fermion mode.
This result provides a clear  and consistent interpretation
of the absorptive part of the self-energy in terms of
a single physical quantity.
Further, since the damping rate is expressed in terms of transition
probabilities, our work could serve to consider the problem of
the gauge dependence of the fermion damping rate from a different
point of view.

\begin{center}
Acknowledgement
\end{center}
This work was partially supported by Grant  DGAPA-IN100691
at the Universidad Nacional Aut\'onoma de M\'exico,
and by the US National Science Foundation Grant PHY-9320692 at
the University of Puerto Rico.

\newpage
\begin{center}{\bf Figure Captions}
\end{center}
\begin{description}
\item[Fig. 1.] Self-energy diagram for the fermion $f$
due to the interaction $\lambda\overline \psi_R f_L\phi$.
\end{description}


\begin{references}
\bibitem{footnote1} With this convention for $\gamma$,
		the \underline{probability} that the particle has not
		decayed behaves as $\sim e^{-\gamma t}$.
\bibitem{footnote3} Since $f_r$ and $f_i$ are defined
		to be functions of the complex variable $\varepsilon$,
		they coincide with the real and imaginary parts of
		$f$ only for real values of that variable, but not otherwise.
\end{references}
\end{document}